\documentclass[manuscript]{aastex}
\usepackage{shortvrb}
\usepackage{graphicx}                    
\usepackage{amssymb}                    
\usepackage{color}                       
\usepackage{url}                         

\begin{document}

\title{Polar Network Index as a magnetic proxy for the solar cycle studies }

\author{Muthu Priyal$^{1}$, Dipankar Banerjee$^{1}$, Bidya Binay Karak$^{2}$, Andr\'es Mu\~noz-Jaramillo$^{3}$, B.~Ravindra$^{1}$, 
 Arnab Rai Choudhuri$^{4}$, Jagdev Singh$^{1}$}

\affil{$^{1}$Indian Institute of Astrophysics,Koramangala, Bengaluru 560034,\\
       $^{2}$Nordita, KTH Royal Institute of Technology and Stockholm University, Sweden,\\
       $^{3}$Montana State University, Bozeman, MT 59717,\\
       $^{4}$Indian Institute of Science, Bangalore, India. 
       \email{mpriya@iiap.res.in, dipu@iiap.res.in}  
             }
\begin{abstract}
 
The Sun has a polar magnetic field which oscillates with the 11-year sunspot cycle. 
This polar magnetic field is an important component of the dynamo process which is operating in the solar convection zone and produces the sunspot cycle. We have systematic direct measurements of the Sun's polar magnetic field only from about mid-1970s. There are, however, indirect proxies which give us information about this field at earlier times. 
The Ca-K spectroheliograms taken in Kodaikanal Solar Observatory during 1904 -- 2007 have now been digitized with the 4k $\times$ 4k 
CCD and have higher resolution ($\sim 0.86$ arcsec) than the other available historical datasets.   From these Ca-K spectroheliograms, we have developed a completely new proxy (Polar Network Index: PNI) for the Sun's polar magnetic field. 
We calculate the PNI from the digitized images using an automated algorithm and  calibrate our measured PNI against the polar field as measured by the Wilcox Solar Observatory for the period of 1976--1990. This calibration allows us to estimate polar fields for the earlier period up to 1904. The dynamo calculations done with this proxy as input data reproduce the Sun's magnetic behavior for the past  century reasonably well. 

\end{abstract}
\section{Introduction}

The interest in the Sun's polar magnetic field has grown tremendously in recent years with the realization that it is primarily this field which gets stretched by the solar wind to fill up the solar system and that this field provides vital clues for predicting the Sun's future magnetic activity \citep{schatten1978, Schatten13}. Although this field was discovered by Babcock \& Babcock (1955), systematic measurements of the polar field have been made at Wilcox Solar Observatory only since 1976. This field oscillates out of phase with the sunspot cycle, becoming strongest during the sunspot minimum \citep{svalgaard2005}. 
The strength of the polar field at the sunspot minimum appears to have a correlation with the strength of the next sunspot cycle \citep{jiang2007,jaramillo2013}, forming the basis for predicting the next cycle. 

There are some proxies known to provide 
 information about the polar field at earlier times. 
For example, the number of bright faculae in the polar regions counted from white-light images of the Sun obtained at Mount Wilson Observatory (MWO) has been found to be a good proxy for the polar field \citep{sheeley1991, jaramillo2012}.  Another proxy is an index $A(t)$
    obtained from H$\alpha$ charts of the Sun \citep{makarov2001}. As we
   shall point out later, there are serious mismatches amongst the different proxies.  Also, the polar faculae number, which seems to be the best proxy for the polar field right now, is counted manually and involves human errors. The faculae counts from MWO and National Astronomical Observatory of Japan (NAOJ) are listed in Table~1 of \cite{li2002}. While the faculae count of cycle 21 is higher than the faculae count of cycle 20 in the data of NAOJ, the opposite is true in the data of MWO. There is certainly a need to look for independent alternative proxies for the polar field. A  proxy computed from an automated algorithm offers the advantage of more reliable and homogeneous results, which is highly desirable for this type of studies. We propose just such a proxy in this paper.

Simple dynamo theory arguments suggest that the polar field must get concentrated into spatially intermittent flux tubes \citep{choudhuri2003}. Hinode found concentrated kilogauss patches in the polar region which are now shown to be correlated with facular bright points at the photospheric level \citep{kaithakkal2013}.  We expect the spatial intermittency of the magnetic field to persist even at the chromospheric level. Images taken in typical chromospheric lines such as Ca-K show the presence of bright cells, usually known as network, while the surrounding black regions are named internetwork. In quiet regions of the photosphere, magnetic fields are known to be concentrated at the network boundaries and appear brighter than the surroundings. We explore if these brightening can be seen in network boundaries at the chromospheric level, particularly at the polar regions of the sun.

In this Letter, 
we show that the  chromospheric  network bright elements  at high latitudes 
seen in the Ca-K spectroheliograms of Kodaikanal 
can be used as a  proxy of the polar field. We call this proxy as Polar network index (PNI). 
From this proxy we deduce the polar magnetic field during 1904 -- 1990.
We finally test whether the past solar cycles are modeled properly on feeding this polar field data 
into a flux transport dynamo model, which is the most widely studied model of the sunspot cycle at present 
\citep[see, recent reviews:][]{charboneau2010,choudhuri2011}.

The next section discusses how we obtain the polar magnetic field from the Kodaikanal Ca-K spectroheliograms.
Then in \S~3 we present a comparison with polar field values obtained from other proxies and discuss a
dynamo simulation feeding our polar field data, with particular attention to hemispheric asymmetry.
Our conclusions are summarized in~\S~4.

\section{Data and Calibration Techniques}

The spatial resolution of the Ca K spectroheliograms taken at Kodaikanal (hereafter KKL) is of about 
2 arcsec  and the exit slit of the spectroheliograph yields a spectral window of 0.5~\AA~ centered at the Ca-K 
line at 3933.67~\AA.  
\cite{ermoli2009} pointed out that the Kodaikanal 
archive hosts the longest homogeneous record, with less variations in spatial resolution. The earlier version of the 8-bit data at kodaikanal
is sufficient to study the plage area having high intensity contrast, but does not provide the required 
photometric accuracy to properly identify the network structures because of the small intensity contrast of 
these features. Therefore, we have designed and developed two digitizer units, using a 1~m labsphere with 
an exit port of 350 mm which provides a stable and uniform source of light with less than 1\% variation
from the center to the edge of the light source. 
The CCD camera of 4k$\times 4$k format with pixel size of 15 $\mu$ square, 
16-bit read out, operating at temperature of $-100^0$C was used to digitize the images. 
The Ca-K network can be clearly seen because of high spatial resolution of digitization (0.86 arcsec). The raw and 
the calibrated data are archived in \url{http://kso.iiap.res.in/data}. 

\begin{figure}
\centerline{\includegraphics[width=0.56\textwidth]{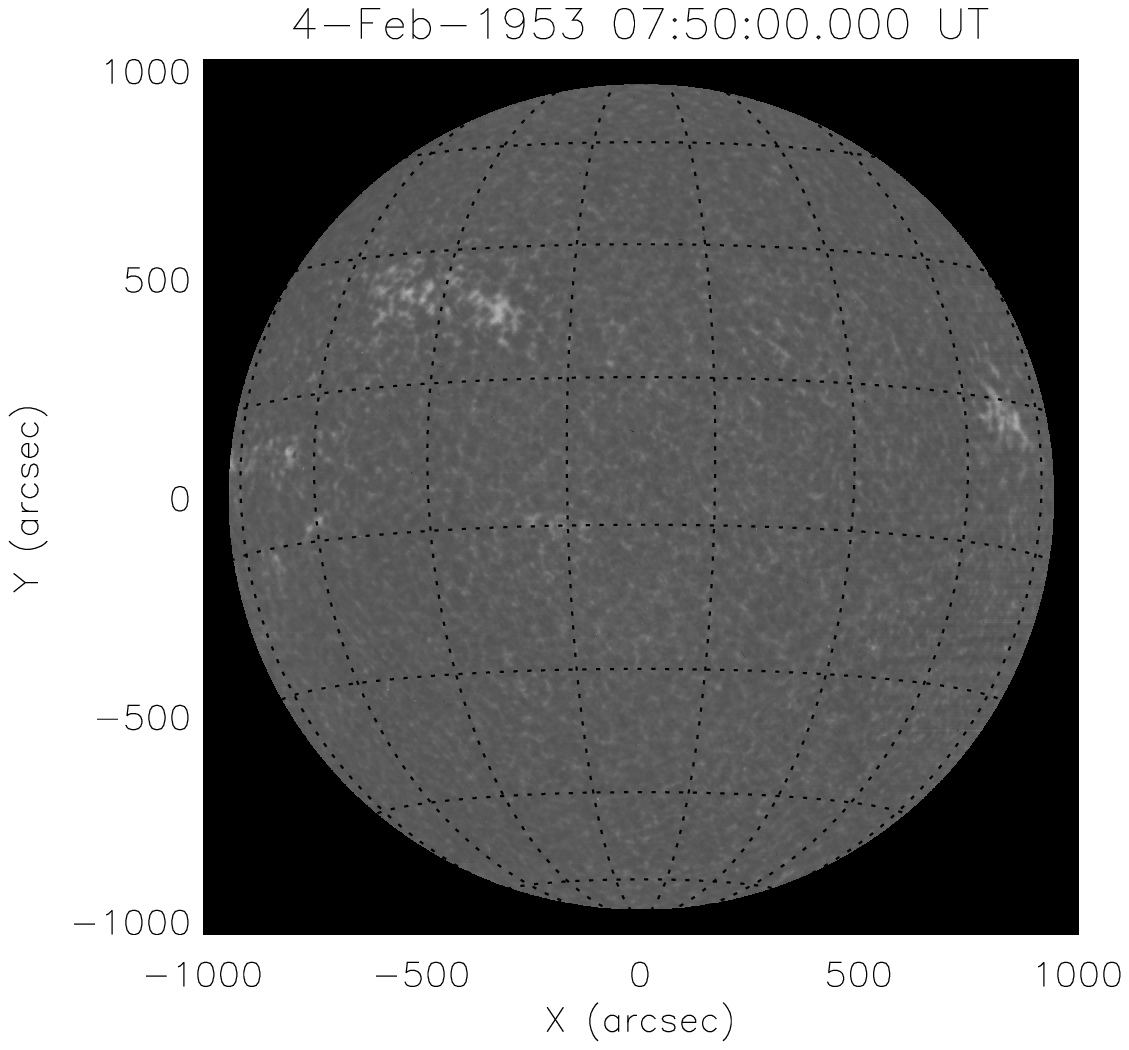}\includegraphics[width=0.56\textwidth]{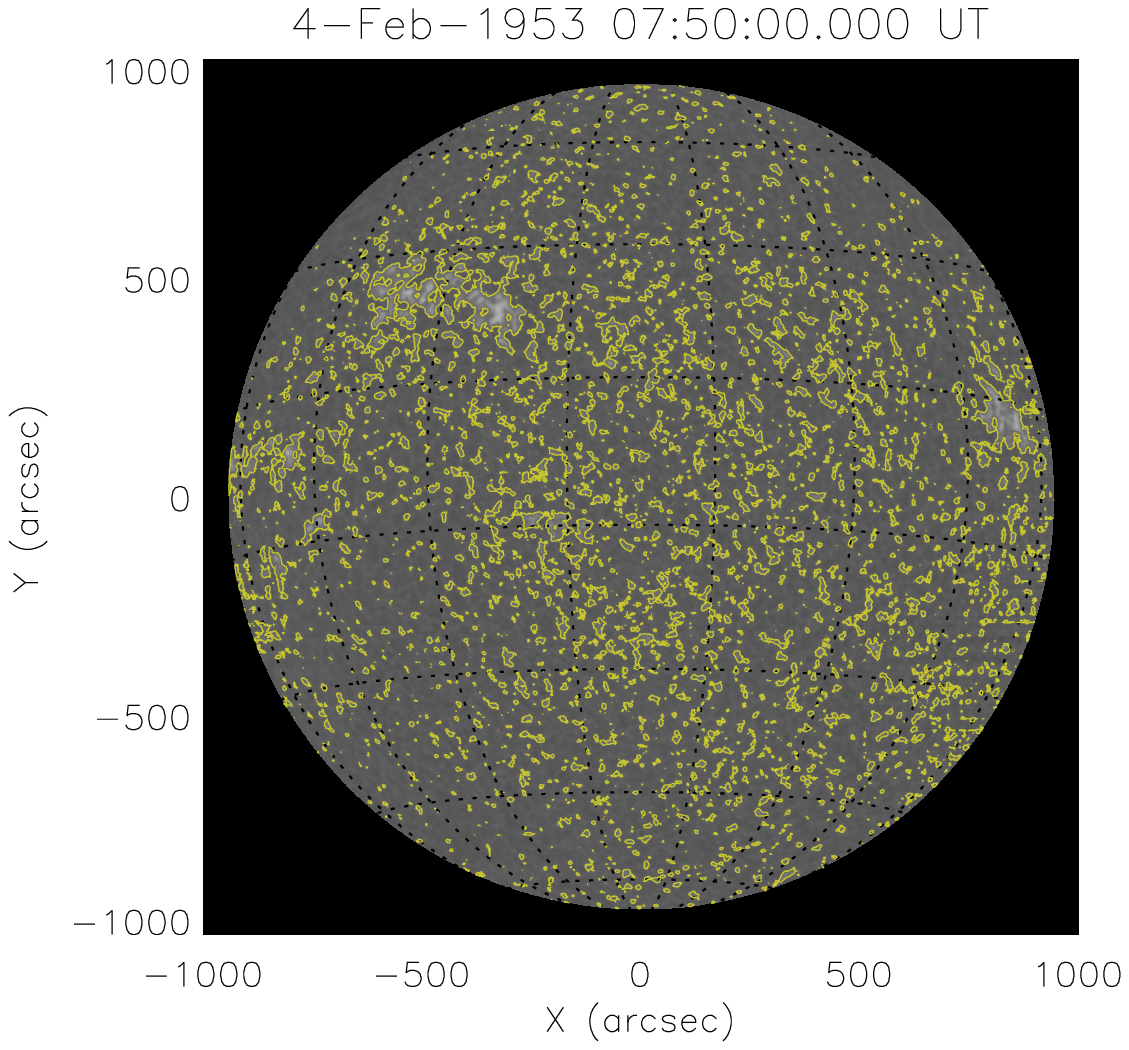}}
\caption{Full disk Ca-II K spectroheliogram (left panel) with contours surrounding features having contrast greater than 1.15 (right panel).}
\label{fig:1}
\end{figure}

The calibration procedure includes density to intensity conversion, flat fielding, limb darkening correction and removal of non-uniformity of the background intensity. After removal of the trend in the intensity profiles, the background of each image becomes uniform. On normalizing the background to unity, the intensity contrast at all locations of the images appears to vary between 1 and 3 depending on the chromospheric features for the data obtained at Kodaikanal. We have identified various chromospheric features by an automated thresholding  technique
\citep{muthupriyal2014}. In this paper we consider  polar regions only and we use a threshold value of 1.15 to identify the small scale magnetic patches. 
Regions above $70^\circ$ 
latitude and with intensity contrast larger than 1.15 have been marked as Polar Network elements. The numbers of these polar network pixels are counted in the latitude belt 70$^\circ$ -- 90$^\circ$ and termed as Polar Network Index (hereafter, PNI). The threshold detection procedure provides the PNI for the Kodaikanal Ca K spectroheliograms for the period 1909--1990.  The filling factor of such pixels in polar regions is  smaller than that in equatorial regions and the intensity contrast is comparable to the equatorial network elements \citep{kaithakkal2013}. Even if one chooses a threshold value slightly different from 1.15, the absolute values change marginally but the relative variations with solar cycle remain the same and the calibration with magnetic field is unaffected. Since polar latitudes above (below) 70$^\circ$ north (south) are best observed during the months of August-September (February-March) we take the average of daily PNI during the two corresponding months for each hemisphere (and averaged over a year). Figure \ref{fig:1} shows a typical spectroheliogram image (left panel) along with features identified with intensity contrast greater than 1.15 (right panel).
A figure at the website \url{http://kso.iiap.res.in/data}  gives the number of available spectroheliogams in different years (in both the northern and southern 
hemispheres) which had been used to construct the yearly PNI.

\section{Results and Discussions}
\subsection{Validation of KKL PNI Index}

\begin{figure}
\centerline{\includegraphics[width=0.9\textwidth]{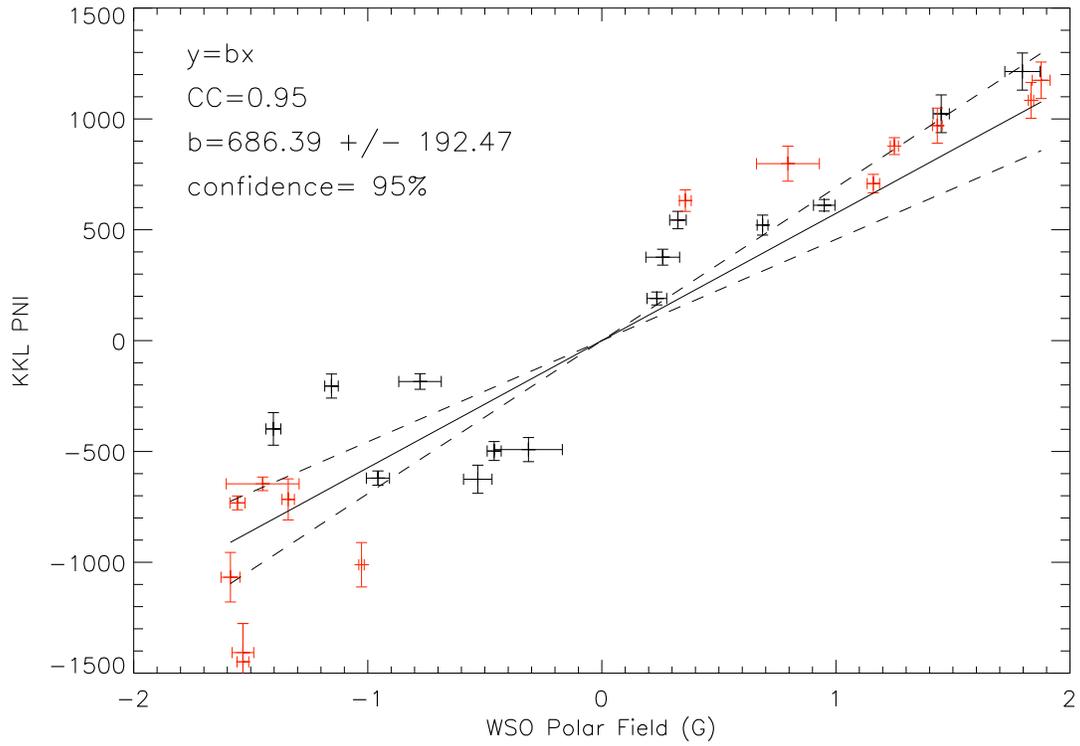}}
\caption{Scatter plot of KKL PNI values with the Wilcox Solar Observatory (WSO) polar field measurements for the period of 1976 -- 1990. Red symbols correspond to the solar minima.}
\label{fig:2}
\end{figure}

We first compare our PNI values with the Wilcox Solar Observatory (WSO) polar field measurements for the period of 
1976 -- 1990 in Figure \ref{fig:2}. We find a good correlation between them (the linear corr. coeff. $\sim 0.95$). The red symbols correspond to the solar minima. It is seen that the polar field tends to be stronger during solar minima. 
The slope $b$ of the linear fit 
provides the calibration to obtain the magnetic field from PNI. 
The calibrated data give the polar fields for the period when no direct polar field measurement exists in any data archive. 
This new estimated polar field has been used as an input to the dynamo simulation. 
To estimate the error bars for the calibrated PNI, we calculate standard deviation ($\sigma$) and 
then errors ($\sigma/\sqrt n$) and compounded it with the 95\% confidence bounds on the 
fitted slope.

\begin{figure} 
\centerline{\includegraphics[width=0.99\textwidth]{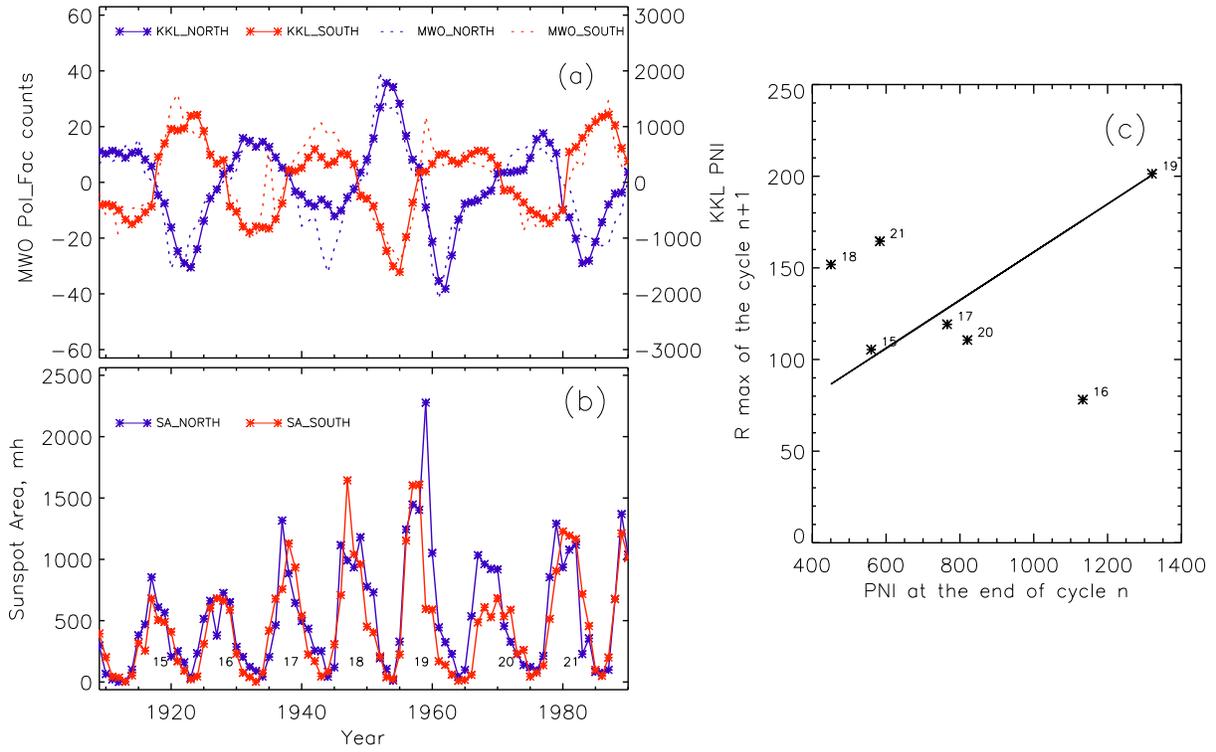}}
\caption{(a) Shows an overplot of PNI and MWO facular count for the years 1909 -- 1990 covering cycles 15 -- 21 for both the northern and southern hemispheres, and (b) shows the sunspot area variations. (c) A correlation plot between the strengths of solar cycles and the PNI values at the preceding minima.}
\label{fig:3}
\end{figure}

\cite{jaramillo2012}  generated polar flux data by validating and calibrating MWO polar faculae counts data against MDI magnetograms and WSO polar field measurements. These data have been taken to validate our KKL PNI data. The yearly averaged data of MWO  Polar faculae include only two months data of each year (Aug-Sep for northern hemisphere and Feb-Mar for the southern hemisphere; \cite{jaramillo2012}) 
in which solar poles are most visible from the Earth. 
Figure \ref{fig:3}(a) shows an overplot of PNI and MWO facular count for the years 1909 -- 1990 covering solar cycles 15--21 for both the northern and southern hemispheres. Though both facular measurements and PNI are strictly positive quantities, they have been given signs (which reverses when they reach a minima) to match the polarity believed to be present at each pole during each solar cycle.  Both indexes are found to be in very good agreement, with a correlation coefficient of 0.91. 
Only around 1945 we notice a significant divergence between them. 
This was a period when there was a dearth of usable spectroheliogram plates to construct PNI (see figure in \url{http://kso.iiap.res.in/data}), probably leading to higher averaging errors. 

The sunspot area data, downloaded from the site \url{http://solarscience.msfc.nasa.gov/greenwch/},
is plotted in Figure \ref{fig:3}(b). A correlation plot of the PNI amplitude at a sunspot minimum and the amplitude of the next sunspot cycle is shown in Figure \ref{fig:3}(c). A similar correlation plot using the MWO facular count was given in Figure 2(b) of \cite{jiang2007}, where two points were outliers. They corresponded to the two weakest cycles 16 and 20 of the last century, which were preceded by not too low facular counts in the preceding minima (around 1923 and 1964 respectively).  However, the $A(t)$ index \citep{makarov2001} was low during these two minima, see Figure 2(a) of \cite{jiang2007}.  We now find the point for cycle 16 to be an outlier when we use PNI for the correlation plot. Although the low value of $A(t)$ during the preceding minimum (around 1923) might suggest a weak polar field, the fact that neither the MWO facular count nor the PNI happened to be too low during this minimum probably indicates that the polar field was not too weak during this minimum.  So we come to the conclusion that at least once in the past century a not too weak polar field during a sunspot minimum was followed by a very weak sunspot cycle, otherwise we find a good correlation between the polar field at the minimum (as indicated by PNI) and the strength of the next sunspot cycle except the points 18 and 21 in Figure~3(c). It may be noted that there were dearth of spectroheliogram plates during the minima preceding cycles 18 and 21 (around 1943 and 1975).  So it is possible that the poor correlations seen for 18 and 21 may be due to errors in the values of PNI.  If we exclude the points 16, 18 and 21, the remaining data points in Figure~3(c) give a correlation coefficient 0.96.
 %

%
 
In addition to the WSO polar field data, the polar faculae counts as recorded from the National Astronomical Observatory of Japan (NAOJ) for the interval of March 1954 to March 1996 have been used for comparison. 
The data have been downloaded from the site \url{http://solarwww.mtk.nao.ac.jp/en/db\_faculae.html}. 
The correlation coefficients between the two datasets are only about 0.65 and 0.33 for the northern and southern hemispheres respectively.
In other words, PNI constructed by us is in good agreement with the MWO faculae counts, but not with the NOAJ faculae counts.
%
%
%
Furthermore, \cite{makarov1989} manually counted  the polar bright points above $50^0$ 
in both the hemispheres for the period 1940--1957 (with a data gap between 1947--1948) 
from KKL Ca-K  spectroheliograms. They found an anti-correlation between polar bright points and 
sunspot number for the two cycles and suggested that Ca-K polar bright points can be used 
as a proxy to predict the characteristics of the next solar cycle. 
We have compared these data with our PNI and
found very good correlation between them.




\subsection{Asymmetry Calculation}

The hemispheric asymmetry of MWO faculae counts at a sunspot minimum is found to have a good correlation 
(correlation coefficient = 0.67)
with the hemispheric asymmetry of the next sunspot cycle (Goel \& Choudhuri 2009). We explore whether the 
hemispheric asymmetry of PNI at a sunspot minimum also has a similar good correlation with the hemispheric 
asymmetry of the next cycle.
	We have computed the hemispheric asymmetry of PNI of the $n$the cycle using the formula
\begin{equation}
\mathrm{ PNI_{AS} = \frac{PNI_N - PNI_S} {PNI_N+PNI_S}},
\end{equation}
where 
$\mathrm{PNI_N}$ and $\mathrm{PNI_S}$ are the north and south PNI during the minimum at the end of the $n$th cycle.
In a similar way, the hemispheric asymmetry of the sunspot area (SA) for the $(n+1)$th cycle is taken as follows
$$\mathrm{SA_{AS} = \frac{SA_N - SA_S} {SA_N+SA_S}}$$
where 
$\mathrm{SA_N}$ and $\mathrm{SA_S}$ are the north and south sunspot area during 
the $(n+1)$th solar cycle. 

\begin{figure} 
\centerline{\includegraphics[width=0.8\textwidth]{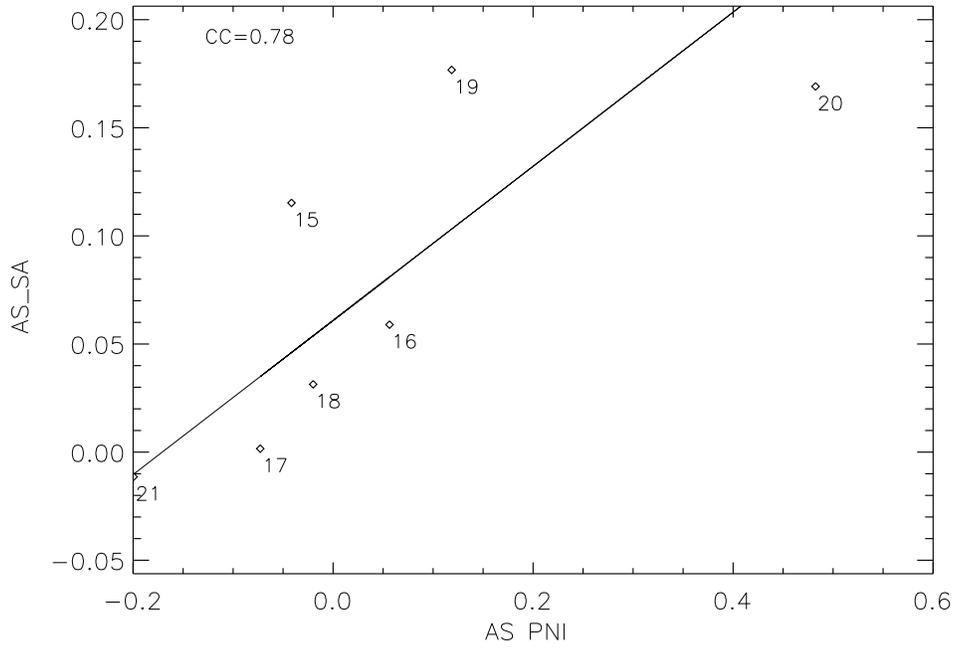}}
\caption{PNI asymmetry of the nth cycle minimum phase is plotted against the sunspot area asymmetry of  the (n+1)th cycle.}
\label{fig:7}
\end{figure}

Figure~ \ref{fig:7}  plots the PNI asymmetry of the $n$th cycle against the SA asymmetry of the $(n+1)$th cycle,
showing a good correlation with a correlation coefficient  around 0.78. 
 We conclude that the hemispheric asymmetry of PNI during a sunspot minimum is a good predictor for the hemispheric asymmetry of the next cycle. 
However, the value of $\mathrm{SA_{AS}}$ for the $(n+1)$th cycle is found to be statistically lower than the value of $\mathrm{PNI_{AS}}$ for the $n$th cycle. A similar result was found for MWO facular counts also \citep{goel2009}. Since both the PNI and the faculae number are proxies for the polar field, we reach the following conclusion. Although the hemispheric asymmetry of the polar field at the sunspot minimum determines (in a statistical sense) the asymmetry of the next cycle, the value of the hemispheric asymmetry of the next cycle turns out to be lower.  During a dynamo cycle, probably the diffusive decay in the solar convection zone tries to reduce the asymmetry introduced in the polar field at the solar minimum \citep{goel2009}.

\subsection{Utilizing the PNI in Flux Transport Dynamo Model}  
In the flux transport dynamo model, the poloidal field is generated by the Babcock-Leighton mechanism. 
It has been proposed that the fluctuations in this mechanism give rise to the irregularities of the sunspot cycle \citep{choudhuri2007, jiang2007}. 
However, of late, we have become aware of a second source of irregularities in the flux transport dynamo: fluctuations in the meridional circulation \citep{karak2010, choudhuri2012} which may complicate things further and may be the reason why the correlation 
between the polar field at the sunspot minimum and the sunspot number of the next cycle is not as good as expected.  
One of the jobs of the flux transport dynamo model is to model the actual sunspot cycles. In order to do this, information about the fluctuations has to be fed to the theoretical model from observational data. 

We now present a simple dynamo calculation to demonstrate that the PNI can be used very conveniently to model the fluctuations in the Babcock-Leighton mechanism. The calculations are based on the code Surya. The early results obtained with this code were presented in \cite{nandy2002}, 
We use the same parameters as used in \cite{KC11} except $v_0$ which is taken here as $25.3$~m~s$^{-1}$ to get correct cycle periods.
If this dynamo code is run uninterrupted, we get periodic solutions in which the polar 
field produced at the end of each cycle would have equal strength. We take the point of view that it 
is the fluctuations of the Babcock-Leighton mechanism which primarily cause variations of the polar 
field from cycle to cycle. In order to model actual cycles, we need to feed the information about the poloidal field in each sunspot minimum. A method for this has been proposed  by \cite{choudhuri2007} 
which was further extended by correcting the poloidal field in two hemispheres separately by \cite{goel2009}, 
where the MWO polar faculae counts of the two poles of the Sun were used. 

Here we use the polar field data as derived from the PNI during solar minimum as a 
measure of the Sun's poloidal field. Due to the limited and noisy nature of the data, 
we calculate the average over 5~yr around the solar minimum to estimate the strength of the polar field 
during that minimum. We compute this value for each cycle and for each hemisphere separately. 
So, corresponding to each cycle, we get one value for N-hemisphere and one for S-hemisphere. 
We rescale these values by the mean of these values to get the strengths of the polar field 
$\gamma_\mathrm N$ = [0.62, 1.41, 0.74, 1.00, 1.63, 1.07, 0.78, 1.30], 
$\gamma_\mathrm S$ = [0.63, 1.21, 0.95, 1.00, 1.41, 0.55, 0.71, 1.26] 
at the ends of cycles 14-21 for northern and southern hemispheres respectively. 
Note that due to insufficient observations in cycle 17 the value of PNI becomes very low 
and this can produce incorrect result. Therefore for this cycle we take $\gamma_\mathrm N$ = $\gamma_\mathrm S = 1.0$, 
i.e., average polar field. We stop the dynamo code at each sunspot minimum and `correct' 
the value of the poloidal field exactly the way it was done in \cite{goel2009}. 
The newly generated poloidal field at the solar minimum remains mostly 
above $0.8R_{\odot}$. Therefore, correcting this poloidal field above $0.8R_{\odot}$ 
by multiplying with $\gamma_\mathrm N$ in the northern hemisphere and with $\gamma_\mathrm S$ 
in the southern hemisphere captures the effect of fluctuations in the Babcock-Leighton process 
to a great extent. After such a correction at a minimum, the dynamo code runs till the next minimum 
when it is stopped again and the poloidal field is corrected. 
From this dynamo calculation we obtain the strength of sunspot activity as a function of time by counting the number of eruptions (assumed to take place whenever the toroidal field above the base of the convection zone exceeds a critical value). Figure~5 gives a plot of the observed sunspot area data for the two hemispheres, along with the theoretically obtained sunspot activity strength suitably scaled to match the sunspot area data.
We notice that the cycles 16 and 21 are not modelled well, 
which we would expect on the ground that the data points 
for these cycles were outliers in Figure \ref{fig:3}(c). 
Most of the other cycles are modelled reasonably well, demonstrating that PNI is indeed a 
useful index for reconstructing the past history of solar activity.

\section{Conclusions}
In recent years there has been considerable effort to develop accurate methods for solar cycle prediction \citep[e.g.,][]{petrovay10}. 
The polar field at the sunspot minimum appears to be the best predictor for the strength of the next cycle.
Since reliable direct measurements of the polar field exist only from 1970s, reconstructing its history at earlier
times has become a major challenge in solar research. 
Our newly digitized Ca-K data at the Kodaikanal archive allow us to identify small features even in the polar region with low contrast.
Using an automated  thresholding method, we have developed a completely new proxy (PNI) for the Sun's polar field and created a  PNI index time series. 
We then combine it with MWO polar flux measurements in order to obtain a 
calibration factor which allows us to estimate the polar fields for a period over which direct 
magnetic field measurements are not available.
Finally we use this newly derived proxy of the polar field as the input 
in a flux transport dynamo model. We find that it reproduces the Sun's magnetic behavior for the past 
 century reasonably well. 
We believe that our data will provide a new tool to understand and model 
the solar magnetic cycles in past.

\clearpage
%
\begin{figure} 
\centerline{\includegraphics[width=1.0\textwidth]{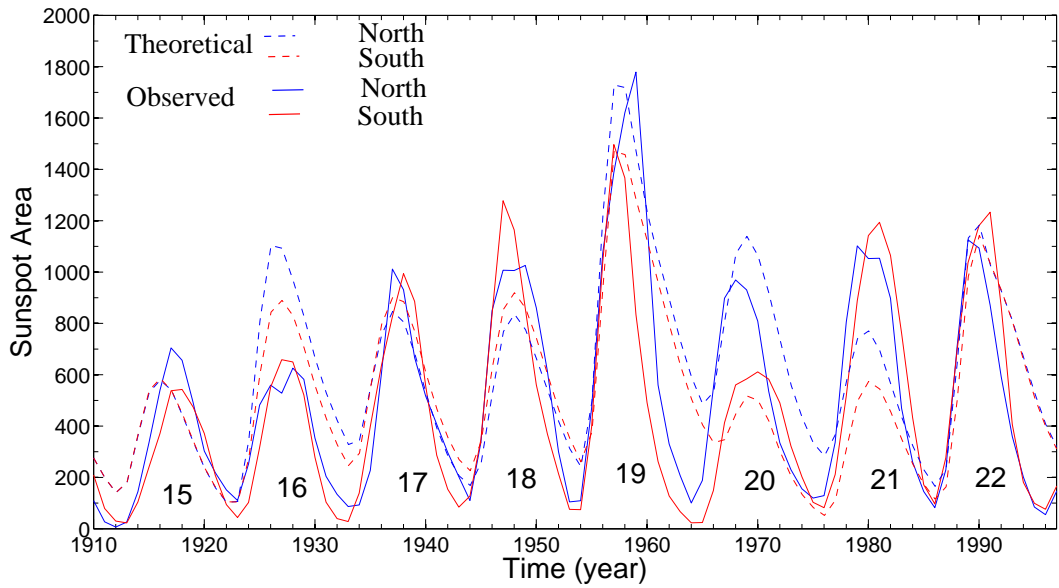}}  
\caption{
The dashed lines show the theoretical sunspot numbers (blue and red corresponding to the northern 
and the southern hemispheres). The theoretical solar cycle is obtained by first considering the 
yearly number of eruptions and then smoothing it. The solid lines show the observed sunspot area 
data (blue for north and red for south). Both the theoretical and observational data are smoothed 
with a Gaussian filter having FWHM = 2~yr.
}
\label{fig:5}
\end{figure}

\end{document}